\documentclass{PoS}

\usepackage{amsmath,amsfonts,amssymb,slashed}

\usepackage{slashed}

\def\one{\mbox{1 \kern-.59em {\rm l}}}

\def\R{\mathbb{R}}

\def\Z{\mathbb{Z}}
\def\C{\mathbb{C}}

\def\@QC#1{\mathpalette{\setbox0=\hbox\bgroup$\rm}%
  {\egroup C$\egroup\rm\rlap{\kern0.4\wd0\vrule
  width 0.05\wd0 height 0.97\ht0 depth -0.01\ht0}%
  #1\bgroup}}

\def\cN{{\cal N}}
\def\cM{{\cal M}}
\def\cO{{\cal O}}
\def\cC{{\cal C}}
\def\cJ{{\cal J}}

\newcommand{\msu}{\mathfrak{s}\mathfrak{u}}

\def\nn{\nonumber}
\def\bea{\begin{eqnarray}}
\def\eea{\end{eqnarray}}
\def\be{\begin{equation}}
\def\ee{\end{equation}}

\newcommand{\eq}[1]{(\ref{#1})}
\def\Tr{{\rm Tr}}

\def\a{\alpha}          
\def\b{\beta}           
\def\d{\delta}

 \def\L{\Lambda} 

\renewcommand{\L}{\Lambda}

\def\cA{{\cal A}}  \def\cC{{\cal C}}
  
 \def\cH{{\cal H}} \def\cI{{\cal I}}
\def\cJ{{\cal J}} \def\cK{{\cal K}} \def\cL{{\cal L}}
\def\cM{{\cal M}} \def\cN{{\cal N}} \def\cO{{\cal O}}

\def\LNC{\L_{\rm{NC}}}
\def\Mat{\rm{Mat}}
\def\del{\partial}
\def\bDelta{{\bf\Delta}}
 \def\bDelta{{\Box}}

\title{Non-commutative geometry and matrix models}

\ShortTitle{Non-commutative geometry and matrix models}

\author{\speaker{Harold Steinacker}\\
        Fakult\"at f\"ur Physik, Universit\"at Wien \\
Boltzmanngasse 5, A-1090 Wien, Austria\\
        E-mail: \email{harold.steinacker@univie.ac.at}}

\abstract{These notes provide an introduction to the noncommutative  matrix geometry 
which arises within matrix models of Yang-Mills type. 
Starting from basic examples of compact fuzzy spaces, a general notion of embedded noncommutative spaces 
(branes) is formulated, and their effective Riemannian geometry is elaborated. 
This class of configurations is preserved under small deformations, and is therefore
appropriate for matrix models. 
A realization of  generic 4-dimensional geometries is sketched, and
the relation with spectral geometry and with noncommutative 
gauge theory is explained.
In a second part, dynamical aspects of these matrix geometries are considered. 
The one-loop effective action for the  maximally supersymmetric IKKT or IIB matrix model
is discussed, which is  well-behaved on 4-dimensional branes.}

\FullConference{3rd Quantum Gravity and Quantum Geometry School \\
		 February 28 - March 13, 2011\\
		 Zakopane, Poland}

\begin{document}

\section{Introduction}

Our basic notion of space and time go back to Einstein. Space-time is 
described in terms of a pseudo-Riemannian manifold, whose dynamical metric describes gravity through the  
Einstein equations. This concept also provides a basis for quantum field theory, 
where  the metric is usually assumed to be flat,
focusing on the short-distance aspects of the fields living on space-time.

Despite the great success of both general relativity and quantum field theory, there are 
good reasons why we should question these classical notions of space and time.
The basic reason is that nature is governed by quantum mechanics. Quantum mechanics is fundamentally different from 
classical physics, and the superposition principle rules out a description in terms of sharply defined classical 
objects and states. Since general relativity (GR) couples matter with geometry, a superposition of matter
 entails also a superposition of geometries. We are thus forced to look for a 
consistent quantum theory of fields {\em and} geometry, hence of gravity. 

Constructing a quantum theory of gravity is clearly a  difficult task. 
Not only is general relativity not renormalizable, 
there are  arguments which suggest that the classical geometric concepts are inappropriate
at very short distances. 
A simple folklore argument goes as follows:  localizing an object at a scale 
$\Delta x$ in quantum mechanics requires to invoke wave-numbers $k \sim \frac 1{\Delta x}$, and thus 
an energy of order $E = \hbar k \sim \frac{\hbar}{\Delta x}$. 
Now in general relativity, a localized energy $E$ defines a length scale given by the
corresponding Schwarzschild radius $R_{\rm Schwarzschild} \sim G E \geq \frac{\hbar G}{\Delta x}$.
Since observations inside trapped surfaces do not make sense,
one should require $(\Delta x) \geq R_{\rm Schwarzschild} \geq \frac{\hbar G}{\Delta x}$, hence
$(\Delta x)^2 \geq \hbar G = L_{Pl}^2$. 
Of course the argument is over-simplistic, 
however a refined argument \cite{Doplicher:1994tu} suggests that quantum mechanics combined with
GR implies uncertainly relations for the space-time coordinates at the Planck scale. 
Even if one does not want to take such ``derivations'' too serious, there is  common consensus that 
space-time should become fuzzy or foam-like at the Planck scale. Canonical or loop quantum gravity 
indeed leads to an area quantization at the Planck scale, and in string theory something similar
is expected to happen \cite{Amati:1988tn}.

The  short-distance aspects of space-time are problematic 
also within quantum field theory (QFT), leading to the well-known UV divergences. 
They can be handled in renormalizable QFT's, but imply that some low-energy properties of the 
models are very sensitive to the
short-distance physics. This leads to serious fine-tuning problems e.g. for 
the mass of scalar (Higgs!) fields, which strongly suggests new physics at short distances
unless one is willing to accept an anthropic point of view.
Taking into account also gravity leads to even more serious fine-tuning problems,
notably the notorious cosmological constant problem.
The  point is that vacuum fluctuations in quantum field theory couple to the 
background metric, which leads to an induced gravitational action, in particular to  
an induced cosmological constant. Lacking any natural subtraction scheme for these terms, these 
contributions are strongly UV divergent, or very sensitive to
the UV details of the model. No convincing solution to this 
problem has been found, which should arise in all approaches based on general relativity including loop
quantum gravity and string theory.

Given all these difficulties, we will discuss a radically different approach here. 
Since the notions of space-time and geometry were argued to make sense only at macroscopic
scales,  the basic degrees of freedom in a fundamental quantum theory
may be very different from the macroscopic ones, while space-time and geometry  ``emerge'' 
in some semi-classical sense. This idea is of course not new, and there are 
many models where some effective metric emerges in composite systems. 
However, we need a model which leads to a {\em universal} dynamical metric
coupling to a (near-) realistic quantum field theory, 
simple enough to admit an analytic understanding. 

In these notes, we will discuss certain  specific matrix models  of Yang-Mills type, which 
seem to realize this idea of emergent geometry and gravity in a remarkably simple way.
These models have been put forward in string theory \cite{Ishibashi:1996xs,Banks:1996vh}, 
and may provide a description 
for the quantum structure of space-time and geometry. 
The beauty lies in the simplicity of these models, whose structure  is 
\be
S_{YM} = {\rm Tr} [X^a,X^b][X^{a'},X^{b'}] g_{aa'} g_{bb'} \,\, + {\rm fermions} .
\label{MM-action}
\ee
Here $X^a,\, a= 1,...,D$ are a set of hermitian matrices, and we focus on the case of Euclidean signature 
$g_{ab} = \d_{ab}$  in this article. No notion of differential geometry 
or  space-time  is used in this action. 
These geometrical structures arise in terms of solutions and fluctuations of the models. 
The aim of this article is to clarify the scope and the mathematical description of this matrix geometry.

Simple examples of such matrix geometries, notably the fuzzy sphere $S^2_N$ or more general quantized homogeneous
spaces including the Moyal-Weyl quantum plane $\R^{2n}_\theta$, have been studied in great detail.
However to describe the general geometries required for gravity, one cannot rely on their special group-theoretical 
structures. 
The key is to consider generic quantizations of submanifolds or 
{\em embedded noncommutative (NC) branes}  $\cM \subset \R^D$ in Yang-Mills matrix models
\cite{Steinacker:2010rh}. 
This provides a sufficiently large class of
matrix geometries to describe realistic space-times. 
Their effective geometry is easy to understand in the ''semi-classical limit``, 
where commutators are replaced by Poisson brackets. 
$\cM$ then inherits the pull-back metric $g_{\mu\nu}$ of $\R^D$, which combines 
with the Poisson (or symplectic) structure $\theta^{\mu\nu}(x)$ to form an effective metric $G_{\mu\nu}(x)$. 
Our task is then to elaborate the resulting physics of these models, and to 
identify the necessary mathematical structures to understand them.


The aim of these notes is to provide a basic understanding of 
matrix geometry and its mathematical description, and to explain the physical relevance of matrix models.
We first recall in detail some examples of matrix geometries with special symmetries.  
This includes well-known 
examples such as the fuzzy sphere, fuzzy tori, cylinders, and the quantum plane. 
We then explain how to extract the geometry without relying on particular symmetries.
The spectral geometry of the canonical  Laplace operator is discussed, and compared with a 
semi-classical analysis. An effort is made to illustrate the scope and generality of 
matrix geometry. The remarkable relation between matrix geometry and
noncommutative gauge theory \cite{Rivelles:2002ez,Yang:2004vd,Steinacker:2007dq} is also discussed briefly.

Our focus on matrix geometry is justified by the good behavior of certain Yang-Mills matrix models -- more precisely, of
one preferred incarnation given by the IKKT model \cite{Ishibashi:1996xs} -- under quantization.
This will be explained in section \ref{sec:quantization}.
The IKKT model is singled out by supersymmetry and its (conjectured) UV finiteness on 4-dimensional backgrounds, 
and it may provide just the right degrees of freedom for a quantum theory of fundamental interactions.
All the ingredients required for physics may emerge from the model, and 
there is no need to add additional structure.
Our strategy is hence to study the resulting physics of these models and  to
identify the appropriate structures, while minimizing any mathematical assumptions or prejudices.
It appears that Poisson or symplectic structures do play a central role.
This is  the reason why the approach presented here does not follow
Connes axioms \cite{connes} for noncommutative geometry, but we will indicate some 
relations  where appropriate.

\section{Poisson manifolds and quantization}

We start by recalling the concept of the quantization 
of a Poisson manifold $(\cM,\{.,.\})$, referring e.g. to 
\cite{Bordemann:1993zv} and references therein 
for more mathematical background. 
A Poisson structure is an anti-symmetric bracket 
$\{.,.\}:\,\cC(\cM)\times \cC(\cM) \to \cC(\cM)$ which 
is a derivation in each argument and satisfies the Jacobi identity,
\be
\{fg,h\} = f\{g,h\} + g\{f,h\}, \qquad \{f,\{g,h\}\} + {\rm cycl.} = 0 .
\ee
We will usually assume that $\theta^{\mu\nu} = \{x^\mu,x^\nu\} $
is non-degenerate, thus defining a symplectic form
\be
\omega = \frac 12 \theta^{-1}_{\mu\nu} dx^\mu dx^\nu 
\ee
in  local coordinates.
In particular, the dimension $\dim \cM = 2n$ must then be even, and
$d\omega = 0$ is equivalent to the Jacobi identity.

It is sometimes useful to introduce 
an expansion parameter $\theta$ of dimension ${length}^2$ and write
\be
\{x^\mu,x^\nu\} = \theta^{\mu\nu}(x) = \theta\,\theta_0^{\mu\nu}(x)
\ee
where $\theta_0^{\mu\nu}$ 
is some fixed Poisson structure.
Given a Poisson manifold, we denote as
{\em quantization map} an isomorphism of vector spaces
\be
\begin{array}{rcl}
\cI: \quad \cC(\cM) &\to& \cA\,\,\subset \,\, \Mat(\infty,\C)\, \\
 f(x) &\mapsto& F
\end{array}
\label{quant-map} 
\ee
which depends on the Poisson structure $\cI \equiv \cI_\theta$, 
and satisfies\footnote{The precise definition of this
limiting process is non-trivial, and there
are various definitions and approaches. 
Here we simply
assume that the limit and the expansion in $\theta$ exist
in some appropriate sense.}
\be
\cI(f g) - \cI(f)\cI(g) \,\, \to \,\, 0 \quad\mbox{and}\quad
\frac 1\theta \Big(\cI(i\{f,g\}) - [\cI(f),\cI(g)]\Big) \,\, \to \,\, 0 
\qquad \mbox{as}\quad \theta \to 0 .
\label{poisson-comp}
\ee
Here $\cC(\cM)$ denotes a suitable space of functions on $\cM$, 
and $\cA$ is interpreted as quantized algebra of 
functions\footnote{$\cA$ is the algebra 
generated by $X^\mu = \cI(x^\mu)$, or some 
subalgebra corresponding to well-behaved functions.} 
 on $\cM$. Such a quantization map $\cI$ is not unique, 
i.e. the higher-order terms in \eq{poisson-comp} are not unique.
Sometimes we will only require that $\cI$ is injective after 
a UV-truncation to
$\cC_\L(\cM)$ defined in terms of a Laplace operator, where $\L$ is some UV cutoff. 
This is sufficient for physical purposes.
In any case, it is clear that $\theta^{\mu\nu}$ -- if it exists in nature --  
must be part of the dynamics of space-time. This will be discussed below.

The map $\cI$ allows to define a 
``star'' product on $\cC(\cM)$ as the pull-back of the algebra $\cA$,
\be
f \star g := \cI^{-1}(\cI(f) \cI(g)) .
\label{star-product}
\ee 
It allows to work with classical functions, 
hiding $\theta^{\mu\nu}$ in the star product.
Kontsevich has shown \cite{Kontsevich:1997vb} that such a quantization 
always exists in the sense of formal power series in $\theta$.
This is a bit too weak for the present context since 
we deal with operator or matrix quantizations. In the case of 
compact symplectic spaces, existence proofs for quantization maps
in the sense of operators as required here are available \cite{Bordemann:1993zv}, and we will not worry about
this any more.
Finally, the integral over the classical space is related in the semi-classical limit
to the trace over its quantization as follows
\be
\int\frac{\omega^n}{n!} f \,\sim \,  (2\pi)^n Tr \cI(f) .
\label{int-trace}
\ee

\paragraph{Embedded noncommutative spaces.}

Now consider  a Poisson manifold embedded in $\R^D$. 
Denoting the Cartesian coordinate functions on $\R^D$ with $x^a,\, a= 1,...,D$, the embedding is 
encoded in the maps
\be
x^a:\quad \cM \hookrightarrow \R^D ,
\ee
so that $x^a \in \cC(\cM)$. Given a quantization \eq{quant-map} of the Poisson manifold $(\cM,\{.,.\})$, we obtain 
quantized embedding functions 
\be
X^a := \cI(x^a) \,\in\, \cA\,\subset \,\Mat(\infty,\C)
\ee
given by specific (possibly infinite-dimensional) matrices. 
This defines an {\em embedded noncommutative space}, or a NC brane. These provide a natural 
class of configurations or backgrounds for the matrix model \eq{MM-action}, 
which sets the stage for the following considerations.
The map \eq{quant-map} then allows to identify 
elements of the matrix algebra with functions on the classical space,
and conversely the commutative space arise as a useful approximation of some matrix background.
Its Riemannian structure will be identified later.

Note that given some arbitrary matrices $X^a$, there is in general no classical space for which this interpretation
makes sense. Nevertheless, we will argue below that this class of backgrounds
is in a sense stable and preferred by the matrix model action, and
this concepts seems appropriate to understand the physical content of the matrix models
under consideration here.

Let us discuss the {\em semi-classical limit} of a noncommutative space. In practical terms,
this means that every matrix $F$ will be replaced by
its classical pre-image $\cI^{-1}(F) =: f$, and commutators will be replaced
by Poisson brackets. The semi-classical limit provides the leading classical approximation 
of the noncommutative geometry, and will be denoted as $F \sim f$ .
However one can go beyond this semi-classical limit using e.g. the star product, which
allows to systematically interpret the NC structure in the language of classical functions and geometry,
as higher-order corrections in $\theta$ to the semi-classical limit. 
The matrix model action \eq{MM-action} can then be considered as 
a deformed action on some underlying classical space.
This approximation is useful if the higher-order corrections in $\theta$ are small.

\section{Examples of matrix geometries}
\label{sec:examples}

In this section we discuss some basic examples of embedded noncommutative spaces
described by finite or infinite matrix algebras. The salient feature is that the geometry is defined by a specific set of 
matrices $X^a$, interpreted as quantized embedding maps of a sub-manifold in $\R^D$.

\subsection{Prototype: the fuzzy sphere}

The fuzzy sphere  $S^2_N$  \cite{Madore:1991bw,hoppe} is a 
quantization or matrix approximation of the usual sphere $S^2$,
with a cutoff in the angular momentum.
We first note that the algebra of functions on the ordinary
sphere can be generated by the coordinate functions $x^a$ of $\R^3$ modulo the
relation $ \sum_{ {a}=1}^{3} {x}^{a}{x}^{a} = 1$. 
The fuzzy sphere $S^2_{N}$ is a
non-commutative space defined in terms of three $N \times N$ hermitian matrices $X^a, a=1,2,3$ 
subject to the relations
\begin{equation}
[ X^{{a}}, X^{{b}} ] = \frac{i}{\sqrt{C_N}}\varepsilon^{abc}\, X^{{c}}~ , 
\qquad \sum_{{a}=1}^{3} X^{{a}} X^{{a}} =  \one 
\end{equation}
where $C_N= \frac 14(N^2-1)$ is the value of the quadratic Casimir of $\msu(2)$ on $\C^N$.
They are realized by the generators of the $N$-dimensional representation $(N)$ of $\msu(2)$. 
The matrices $X^a$ should be interpreted as quantized embedding functions
in the Euclidean space $\R^3$,
\be
X^a \sim x^a:\quad S^2 \hookrightarrow \R^3.
\label{embedding-S2}
\ee
They generate an algebra $\cA \cong \Mat(N,\C)$,
which should be viewed as quantized algebra of functions on the symplectic space $(S^2,\omega_N)$
where $\omega_N$ is the canonical $SU(2)$-invariant symplectic form on $S^2$
with $\int \omega_N = 2\pi N$. 
The best way to see this is to decompose $\cA$ into irreducible representations under the adjoint action of $SU(2)$, 
which is obtained from 
\bea
S^2_{N} \cong (N) \otimes (\bar N) 
&=& (1) \oplus (3) \oplus ... \oplus (2N-1) \nn\\
&=& \{\hat Y^{0}_0\} \,\oplus \, ... \, \oplus\, \{\hat Y^{N-1}_m\}.
\label{fuzzyharmonics}
\eea
This provides the definition of the 
fuzzy spherical harmonics $\hat Y^{l}_m$, and defines the {\em quantization map}
\be
\begin{array}{rcl}
\cI: \quad \cC(S^2) &\to& \cA\,\,= \,\, \Mat(N,\C)\, \\
 Y^l_m &\mapsto& \left\{\begin{array}{c}
                         \hat Y^l_m, \quad l<N \\ 0, \quad l \geq N
                        \end{array}\right.

\end{array}
\label{quant-map-S^2} 
\ee
It follows easily that $\cI(i\{x^a,x^b\}) = [X^a,X^b]$ where $\{,\}$ denotes the Poisson brackets 
corresponding to the symplectic form $\omega_N = \frac N2 \varepsilon_{abc} x^a dx^b dx^c$
on $S^2$.
Together with the fact that $\cI(f g) \to \cI(f)\cI(g)$ for $N \to \infty$ (which is not hard to prove), 
$\cI(i\{f,g\}) \stackrel{N \to \infty}{\to} [\cI(f),\cI(g)]$ follows. 
This means that $S^2_N$  is the quantization of $(S^2,\omega_N)$.
It is also easy to see the following integral relation
\be
2\pi\, \Tr (\cI(f)) = \int\limits_{S^2} \omega_N f,
\ee
consistent with \eq{int-trace}.
Therefore $S^2_N$ is  the  quantization of $(S^2,\omega_N)$. 
Moreover, there is a natural Laplace operator\footnote{The symbol $\Box$ is used here to distinguish
the matrix Laplace operator from the 
Laplacian $\Delta$ on some Riemannian manifold. It does not indicate any particular signature.} on $S^2_N$ defined as
\be
{\Box} = [X^a,[X^b,.]]\d_{ab}
\label{matrix-laplacian-S2}
\ee
which is invariant under $SU(2)$; in fact it is nothing but the 
quadratic Casimir of the $SU(2)$ action on $S^2$. It is then easy to see that up to normalization,  
its spectrum coincides with the spectrum of the classical 
Laplace operator on $S^2$ up to the cutoff, and the eigenvectors are given by the fuzzy spherical harmonics 
$\hat Y^l_m$.

In this special example, \eq{fuzzyharmonics} allows to construct a  series of embeddings of vector spaces
\be
\cA_N \subset \cA_{N+1} \subset ...   
\label{fuzzy-sequence}
\ee
with norm-preserving embedding maps. 
This allows to recover the classical sphere by taking the inductive limit.
While this is a very nice structure, we do not want to rely on the existence of such 
explicit series of embeddings. We emphasize that even finite-dimensional 
matrices allow to approximate a classical geometry to a high precision, as
 discussed further in  section \ref{sec:lessons}.

\subsection{The fuzzy torus}

The fuzzy torus $T^2_N$ can be defined  in terms of clock and shift operators 
$U,V$ acting on $\C^N$ with relations $U V = q V U$
for $q^N=1$, with $U^N = V^N = 1$. They have the following standard representation
\be
 U = \begin{pmatrix}
     0 & 1  & 0 &  ... & 0 \\
     0 & 0 & 1 & \, ...& 0 \\
     && \ddots && \\
     0 &  & ... &  0 & 1 \\
     1 & 0  &   ... &  & 0
    \end{pmatrix}, \quad V =  \begin{pmatrix}
   1 &  &  &   \\
     &  e^{2\pi i\frac 1N}  &     \\
      &  & e^{2\pi i \frac 2N}   \\
    && \ddots & \\
      &  &   \, & e^{2\pi i \frac{N-1}N}
    \end{pmatrix} .
\ee
These matrices generate 
the algebra $\cA \cong \Mat(N,\C)$, which can be viewed as quantization of the function algebra $\cC(T^2)$
on the symplectic space $(T^2,\omega_N)$. 
One way to recognize the structure of a torus is by identifying a
$\Z_N \times \Z_N$ symmetry, defined as 
\begin{align}
\Z_N \times \cA &\to \cA  \hspace{3cm}   \\  
(\omega^k,\phi) &\mapsto U^{k} \phi U^{-k}
\end{align}
and similarly for the other $\Z_N$ defined by conjugation with $V$. 
Under this action, the algebra of functions $\cA = \Mat(N,\C)$ decomposes as
\be
\cA = \oplus_{n,m=0}^{N-1}\, U^n V^m
\ee 
into harmonics i.e. irreducible representations. This suggests to define the following quantization 
map: 
\begin{align}
\cI: \quad \cC(T^2) &\to \cA\,\,= \,\, \Mat(N,\C)\, \qquad\\
 e^{in\varphi} e^{i m \psi} &\mapsto \left\{\begin{array}{c}
                        q^{-nm/2}\, U^n V^m, \quad |n|,|m|<N/2 \\ 0, \quad \mbox{otherwise}
                        \end{array}\right. \nn
\end{align}
which is compatible with the $\Z_N \times \Z_N$ symmetry and satisfies $\cI(f^*) = \cI(f)^\dagger$.
The underlying Poisson structure on $T^2$ is given by
$\{e^{i\varphi},e^{i\psi}\} = \frac{2\pi}N\,e^{i\varphi}e^{i\psi}$ (or equivalently $\{\varphi,\psi\} = -\frac{2\pi}N)$,
and it is easy to verify the following integral relation
\be
2\pi\, \Tr (\cI(f)) = \int\limits_{T^2} \omega_N f,\qquad 
  \omega_N =  \frac{N}{2\pi}\, d\varphi d\psi  
\ee
consistent with \eq{int-trace}.
Therefore $T^2_N$ is  the  quantization of $(T^2,\omega_N)$. 

The metric is an additional structure which goes beyond the mere concept of quantization.
Here we obtain it by considering $T^2$ as embedded noncommutative space in $\R^4$, 
by defining  4 hermitian matrices
\be
X^1 + i X^2 := U, \qquad X^3 + i X^4 := V
\label{torus-embedding}
\ee
which satisfy the relations 
\begin{align}
(X^1)^2 + (X^2)^2 &= 1 = (X^3)^2 + (X^4)^2 ,  \nn\\
(X^1 + i X^2) (X^3 + i X^4) &= q (X^3 + i X^4)(X^1 + i X^2) .
\label{fuzzy-torus}
\end{align}
They can again be viewed as embedding maps
\be
X^a \sim x^a: \quad T^2 \hookrightarrow \R^{4} 
\ee 
and we can write $x^1+ix^2 = e^{i\varphi}, \,\, x^3 + ix^4 = e^{i\psi}$  in the semi-classical limit.
This allows to consider the matrix Laplace operator \eq{matrix-laplacian-S2}, and to compute its spectrum:
\begin{align}
\Box\phi &= [X^a,[X^b,\phi]]\d_{ab} \\[1ex]
 &= [U,[U^{\dagger},\phi]] + [V,[V^{\dagger},\phi]] = 4\phi - U\phi U^\dagger - U^\dagger\phi U 
 -  V\phi V^\dagger - V^\dagger\phi V    \\[1ex]
\Box(U^n V^m) &= c ([n]_q^2 + [m]_q^2)\, U^n V^m \,\sim\, c(n^2+m^2)\, U^n V^m, \nn\\
 c &=  -(q^{1/2} - q^{-1/2})^2 = 4\sin^2(\pi/N) \, \sim \  \frac{4\pi^2}{N^2}
\end{align}
where
\be
[n]_q = \frac{q^{n/2} - q^{-n/2}}{q^{1/2} - q^{-1/2}} = \frac{\sin(n\pi/N)}{\sin(\pi/N)} \sim n
\qquad \mbox{(``q-number'')}
\ee
Thus the spectrum of the matrix Laplacian \eq{matrix-laplacian-S2} approximately 
coincides\footnote{It is interesting to note that momentum space is compactified here, 
reflected in the periodicity of $[n]_q$.} 
with the classical case below the cutoff. 
Therefore $T^2_N$ with the embedding defined via the above 
embedding \eq{torus-embedding} has indeed the  geometry
of a torus.

\subsection{Fuzzy $\C P^N$}

A straightforward generalization of the fuzzy sphere leads to the {\em fuzzy complex projective space} 
$\C P^{n}_N$, which is defined in terms of hermitian matrices $X^a$, $a = 1,2,...,n^2+n$  subject to the 
relations 
\be
[ X^{{a}}, X^{{b}} ] = \frac{i}{\sqrt{C_N'}} f^{ab}_c\, X^{{c}}~, \qquad 
 d_{ab}^c  X^{{a}} X^{{b}} = D_N X^c, \qquad X_a X^a =  \one 
\ee
(adopting a sum convention).
Here $f^{ab}_c$ are the structure constants of $\msu(n+1)$, $d^{abc}$ is the totally symmetric invariant tensor,
and $C_N', D_N$ are group-theoretical constants which are not needed here.
These relations are realized by the generators of  $\msu(n+1)$ acting on irreducible representations 
with highest weight $(N,0,...,0)$ or $(0,0,...,N)$, with dimension $d_N$. Again, 
the matrices $X^a$ should be interpreted as quantized embedding functions
in the Euclidean space $\msu(n+1) \cong \R^{n^2+n}$,
\be
X^a \sim x^a: \quad\C P^{n} \hookrightarrow \R^{n^2+n}.
\ee
They generate an algebra $\cA \cong \Mat(d_N,\C)$,
which should be viewed as quantized algebra of functions on 
the symplectic space $(\C P^{n}, N\omega)$
where $\omega$ is the canonical $SU(n)$-invariant symplectic form on $\C P^{n}$.
It is easy to write down a quantization map analogous to \eq{quant-map-S^2}, 
\be
\cI: \quad \cC(\C P^{n}) \to \cA\,
\label{quant-map-CP}
\ee
using the decomposition of $\cA$ into irreducible representations of $\msu(n+1)$. 
Again, there is a natural Laplace operator on $\C P^{n}_N$ defined as in \eq{matrix-laplacian-S2}
whose spectrum coincides with the classical one up to the cutoff.
A similar construction can be given for any coadjoint orbit 
of a compact Lie group.

\subsection{The Moyal-Weyl quantum plane}

The Moyal-Weyl quantum plane $\R^{2n}_\theta$
is defined in terms of $2n$ (infinite-dimensional) hermitian matrices $X^a \in \cL(\cH)$
subject to the relations
\be
[X^\mu,X^\nu] = i \theta^{\mu\nu} \one
\ee
where $\theta^{\mu\nu} = - \theta^{\nu\mu} \in \R$. Here $\cH$ is a separable Hilbert space.
This generates the ($n$-fold tensor product of the) Heisenberg algebra $\cA$ (or some suitable 
refinement of it, ignoring operator-technical subtleties here), 
which can be viewed as quantization of the algebra of functions on $\R^{2n}$
using e.g. the Weyl quantization map
\be
\begin{array}{rcl}
\cI: \quad \cC(\R^{2n}) &\to& \R^{2n}_\theta\,\,\subset \,\, \Mat(\infty,\C)\, \\[1ex]
 e^{i k_\mu x^\mu} &\mapsto& e^{i k_\mu X^\mu} .
\end{array}
\label{MW-map} 
\ee
Since plane waves are  irreducible representations of the translation group,
this map is again defined as intertwiner of the symmetry group
as in the previous examples.  
Of course, the matrices $X^\mu$ should be viewed as quantizations of the classical
coordinate functions $X^\mu \sim x^\mu:\,\, \R^{2n} \to \R^{2n}$.
Similar as in quantum mechanics, it is easy to see that the noncommutative plane waves satisfy the 
Weyl algebra
\be
e^{i k_\mu X^\mu} e^{i p_\mu X^\mu} = e^{\frac i2 \theta^{\mu\nu}k_\mu p_\nu} \, e^{i (k_\mu + p_\mu) X^\mu} 
\ee
It is also easy to obtain an explicit form for the star product defined  by the above quantization map:
it is given by the famous Moyal-Weyl star product, 
\be
(f \star g)(x) = f(x) e^{\frac i2\theta^{\mu\nu}\overleftarrow\del_\mu\overrightarrow\del_\nu} g(x)
\ee
in obvious notation. This gives e.g.
\begin{align}
x^\mu \star x^\nu &= x^\mu x^\nu + \frac i2 \theta^{\mu\nu}  \nn\\
[x^\mu,x^\nu]_\star &= i \theta^{\mu\nu} .
\end{align}
In this example, we note that
\be
[X^\mu,.] =: i \theta^{\mu\nu} \del_\nu
\ee
provides a reasonable definition of partial derivatives in terms of inner derivations on $\cA$, 
provided $\theta^{\mu\nu}$ is non-degenerate (which we will always assume).
This is justified by the observation 
$[X^\mu,e^{i k_\nu X^\nu}] = - \theta^{\mu\nu} k_\nu\, e^{i k_\mu X^\mu}$ together with 
the identification $\cI$. Therefore these
partial derivatives and in particular the matrix Laplacian 
\begin{align}
 \Box &= [X^\mu,[X^\nu,.]]\d_{\mu\nu} = -\theta^{\mu\mu'}\theta^{\nu\nu'}\d_{\mu\nu}\,\del_{\mu'}\del_{\nu'} \equiv 
 - \LNC^{-4}\, G^{\mu\nu}\del_\mu\del_\nu, \nn\\
G^{\mu\nu} &:= \LNC^{4}\theta^{\mu\mu'}\theta^{\nu\nu'}\d_{\mu'\nu'}, \qquad \LNC^4 :=   \sqrt{\det\theta^{-1}_{\mu\nu}}
\end{align}
coincide via $\cI$ with the commutative Laplacian for the metric $G_{\mu\nu}$. 
Therefore $G_{\mu\nu}$ should be considered as {\em effective metric} of $\R^4_\theta$.

The Moyal-Weyl quantum plane differs from our previous examples in one essential way:
the underlying classical space is non-compact. This means that the matrices become
unbounded operators acting on an infinite-dimensional separable Hilbert space. 
The basic difference can be seen from the formula 
\be
(2\pi)^n \Tr F \sim \int_\cM |\theta^{-1}_{\mu\nu}|\, f\ ,
\ee
which together with $|\theta^{-1}_{\mu\nu}|=const$ implies that the trace diverges as
a consequence of the infinite symplectic volume. Locally (i.e. for test-functions $f$ with 
compact support, say), there is no essential
difference between the compact fuzzy spaces described before and the 
Moyal-Weyl quantum plane. This reflects the Darboux theorem, which 
states that  all symplectic spaces of a given dimension are locally equivalent. 
Thus from the point of view of matrix geometry, 
$\R^{2n}_\theta$ is simply a non-compact version of a fuzzy space.

\subsection{The fuzzy cylinder}

Finally, the {\em fuzzy cylinder} $S^1 \times_\xi \R$ is defined by  \cite{Chaichian:1998kp}
\begin{align}
[X^1,X^3] &= i \xi X^2, & [X^2,X^3] &= - i \xi X^1,  \nn\\
 (X^1)^2 + (X^2)^2 &= R^2, & [X^1,X^2] &= 0  .
\end{align}
Defining $U := X^1+iX^2$ and  $U^\dagger := X^1-iX^2$,
this can be stated more transparently as 
\bea
U U ^\dagger &=& U^\dagger U \quad = R^2     \nn\\   
\, [U, X^3] &=& \xi U,  \qquad [U^\dagger, X^3] = - \xi U^\dagger  
 \label{fuzzy-cylinder} 
\eea
This algebra has the following 
irreducible representation\footnote{More general irreducible representations are obtained 
 by a (trivial) constant shift $X^3 \to X^3 + c$.}
\bea
U |n\rangle &=& R   |n + 1\rangle , \qquad   U^\dagger  |n\rangle = R  |n - 1\rangle  \nn\\
X^3 |n\rangle &=& \xi n  |n\rangle,  \qquad \qquad n \in \Z, \,\, \xi \in \R
\label{cylinder-rep}
\eea
on a Hilbert space $\cH$, where $|n\rangle$ form an orthonormal basis. 
We take $\xi \in \R$, since  the $X^i$ are hermitian. Then
the matrices $\{X^1,X^2,X^3\}$ can be interpreted geometrically as quantized embedding functions
\be
\begin{pmatrix}
 X^1 + i X^2 \\ X^3
\end{pmatrix}
 \sim \begin{pmatrix} R e^{i y_3} \\ x^3
\end{pmatrix}: \quad  S^1 \times \R \hookrightarrow \R^3 .
\ee
The quantization map is given by
\begin{align}
\cI: \quad \cC(S^1 \times \R) \ &\to\ S^1 \times_\xi \R\,\,\subset \,\, \Mat(\infty,\C)\, \\[1ex]
 e^{i p x^3} e^{iny^3} \ &\mapsto \ e^{in\xi/2}\,  e^{i p X^3} U^n,
\end{align}
which preserves the obvious $U(1) \times \R$ symmetry.
This defines the fuzzy cylinder $S^1 \times_\xi \R$. 
It is the quantization of $T^* S^1$ with canonical Poisson bracket $\{e^{iy_3},x^3\} = -i\xi e^{i y_3}$, or
 $\{x^3,y^3\} = \xi$ locally. 
Its  geometry can be recognized either using the $U(1) \times \R$ symmetry, or using the 
matrix Laplacian $\Box = [X^a,[X^b,.]] \d_{ab}$  which has the following spectrum
\be
\Box e^{i p X^3} U^n 
 = \Big(4 R^2 \sin^2(p\xi/2) + n^2 \xi^2 \Big)\,  e^{i p X^3} U^n 
 \, \stackrel{p \xi \ll 1}{\sim} \, \big(R^2 p^2 + n^2  \big)\, \xi^2 e^{i p X^3} U^n ,
\label{spectrum-cyl}
\ee
consistent with the classical spectrum for small momenta. Therefore the effective
geometry is  that of a cylinder.

\subsection{Additional structures: coherent states and differential calculus}
\label{sec:coherent-states}

Next we exhibit some additional results in the example of the fuzzy sphere. They can be 
generalized to other matrix geometries under consideration here.

\paragraph{Coherent states.}

As in Quantum Mechanics,  coherent states provide a particularly illuminating way to understand 
quantum geometry, via maximally localized wave-functions. In the case of $S^2_N$ they 
go back to Perelomov \cite{Perelomov:1986tf}, although different approaches are possible \cite{Grosse:1993uq}.
Here we consider the group-theoretical approach of Perelomov. Let $p_0 \in S^2$ be a given point on $S^2$ called north pole.
Then the map
\begin{align}
 SO(3) &\to S^2   \nn\\
 g &\mapsto g\triangleright p_0
\end{align}
defines the stabilizer group $U(1)\subset SO(3)$ of $p_0$, so that $S^2 \cong SU(2)/U(1)$.
Now recall that the algebra of functions on the fuzzy sphere is given by $End(\cH)$, where
$\cH$ is spanned by the angular momentum basis $|m,L\rangle,\,\, m = -L,...,L,\,\, N=2L+1$.
Noting that $X^3 |L,L\rangle = \frac{N-2}{\sqrt{N^2-1}} |L,L\rangle$, we can identify
the highest weight state $|L,L\rangle$ as coherent state localized on the north pole. 
We define more generally
\be
|\psi_g\rangle := \pi_N(g) |L,L\rangle
\ee
where $\pi_N$ denotes the $N$-dimensional irrep of $SO(3)$. Since the stabilizer group $U(1)$ of $p_0$
acts on $|\psi_g\rangle$ via a complex phase, 
the associated one-dimensional projector
\be
\Pi_g = |\psi_g\rangle\langle\psi_g|
\ee
is independent of $U(1)$, so that there is a well-defined map
\begin{align}
SO(3)/U(1)\ \cong\  S^2  &\to \ End(\cH) \\
    p \,\, &\mapsto \ \Pi_g = |\psi_g\rangle\langle\psi_g| \quad =: (4\pi)^{-1} \,\d^{(2)}_N(p-x) .
\end{align}
The notation on the rhs should indicate that these are the optimally localized wave-functions on $S^2_N$.
To proceed, we label coherent states 
related by $U(1)$ with the corresponding point on $S^2$, so that 
$|p\rangle := |\psi_{g(p)}\rangle \sim |\psi_{g(p)'}\rangle$. 
One can then show the following results \cite{Perelomov:1986tf}
\begin{align}
\qquad \int_{S^2} |p\rangle\langle p| &= c \one  \hspace{5cm} \quad \mbox{overcomplete}  \nn\\
|\langle p|p'\rangle| &= \big(\cos(\vartheta/2)\big)^{N-1},  
   \qquad \vartheta=\measuredangle(p,p')\quad \mbox{localization} \,\,p\approx p' \nn \\[1ex]
p_a X^a |p\rangle  &= |p\rangle \nn \\[1ex]
\langle p|X^a|p\rangle &= \Tr X^a \Pi_p \, = \,\, p^a \in S^2 \nn
\end{align}
which provide a justification for the interpretation of $X^a \sim x^a: \ \ S^2 \hookrightarrow \R^3$ 
\eq{embedding-S2} as quantized embedding maps.
One can also show that these states are optimally localized on $S^2$,
\begin{align}
(\Delta X^1)^2 + (\Delta X^2)^2 + (\Delta X^3)^2 & 
  \geq \frac{N-1}{2C_N}\sim \frac 1{2N}
= \sum_a \langle p|(X^a -\langle p|X^a|p\rangle )^2 |p\rangle  .
\end{align}
In principle, such considerations should apply to all matrix geometries under consideration here, although
the explicit realization of such coherent states is in general not known.

%

\paragraph{Differential calculus.}

For many noncommutative spaces with enhanced symmetry, differential calculi have been constructed which 
respect the symmetry. In the case of embedded NC spaces as considered here, this calculus
is typically that of the embedding space $\R^D$, and does not reduce to the standard one
in the commutative limit. Nevertheless it may be quite appealing, and we briefly discuss the example of the 
fuzzy sphere following Madore \cite{Madore:1991bw}. 

A differential calculus on $S^2_N$ is a graded bimodule $\Omega^\star_N = \oplus_{n\geq 0}\, \Omega^n$ over $\Omega^0 = \cA=S^2_N$ 
with an exterior derivative $d:\Omega^n \to \Omega^{n+1}$ compatible with $SO(3)$, which satisfies 
$d^2 = 0$ and the graded Leibniz rule 
$d(\a\b) = d\a\, \b + (-1)^{|\a|} \a\, d\b$. 
Since $[X^a,X^b] \sim i \varepsilon^{abc} X^c$ one necessarily has $dX^a\, X^b \neq X^b d X^a$ and 
$dX^a dX^b \neq -dX^b dX^a$.
It turns out that there is a preferred (radial) one-form which generates the 
exterior derivative\footnote{This formula is modified for higher forms, cf. \cite{Grosse:2000gd}.
$\omega$ should not be confused with the symplectic form.} on $\cA$, 
\be 
d f = [\omega,f] , \qquad\quad \omega = -C_N X^a dX^a .
\ee
It turns out that the calculus necessarily contains 3-forms
\be
\Omega^\star_N = \oplus_{n=0}^3 \Omega^n_N , \qquad\Omega^3_N \ni f_{abc}(X) dX^a dX^b dX^c ,
\ee
which reflects the embedding space $\R^3$.
It turns out that can introduce a frame of one-forms which commute with all functions \cite{Madore:1991bw}:
\be
\xi^a = \omega X^a + \sqrt{C_N}\, \varepsilon^{abc} X^b dX^c , \qquad [f(X),\xi^a] = 0 .
\ee
The most general one-form can then be written as
\be
A = A_a \xi^a \quad\in \Omega^1_N, \qquad\quad A_a \in \cA = \Mat(N,\C) .
\ee
One can define the exterior derivative such that \cite{Grosse:2000gd}
\begin{align}
F &= dA + AA = (Y_a Y_b + i \varepsilon_{abc} Y_c) \xi^a \xi^b \quad\in \Omega^2_N, \\
Y &= \omega + A = Y_a \xi^a = (X_a + A_a) \xi^a\quad\in \Omega^1_N .
\end{align}
These formulae are relevant in the context of gauge theory, encoding the covariant coordinates
$Y^a = X^a + A^a$ which are the basic objects in matrix models. 
Nevertheless the formalism of differential forms may be somewhat misleading, because the one-form 
$Y$ encodes both tangential gauge fields
as well as transversal scalar fields. 
In any case we will not use differential calculi in the following. Our aim is to understand the 
geometrical structures which  emerge from matrix models, without introducing any 
mathematical prejudice. These models  do not require any such additional mathematical structures.

This concludes our brief exhibition of matrix geometries, 
through examples whose geometry was identified using their symmetry properties.
We will learn below how to generalize them for generic geometries,
and how to systematically extract their geometry without using this symmetry. 
In particular, the form of the matrix Laplacian \eq{matrix-laplacian-S2} turns out to be general.
On the other hand, there are also more exotic and singular spaces
that can be modeled by matrices, such as intersecting spaces, stacks of spaces, etc. 
Some well-known NC spaces such as $\kappa$- Minkowski space 
are quantizations of degenerate Poisson structures, which we do not consider
since the effective metric would be degenerate and/or singular.
There are also very different types of noncommutative tori \cite{Connes:1997cr} 
described by infinite-dimensional algebras, reflecting the presence of a non-classical winding sector. 
They are not stable under small deformations, e.g. it is crucial whether $\theta$ is rational or irrational.
In contrast, the embedded NC spaces considered here such as fuzzy tori
are stable under deformations as explained in section \ref{sec:deformations}, and contain no winding modes. 
This seems crucial for supporting a well-defined quantum field theory, 
as discussed in section \ref{sec:quantization}.

Finally we should recall the noncommutative geometry as introduced by A. Connes \cite{connes},
which is based on a Dirac operator $\slashed{D}$ subject to certain axioms. One can  
define a differential calculus based on such a Dirac operator, which is a refined version of
$d \sim [\slashed{D},.]$. The matrix models discussed below indeed provide a Dirac operator
for the matrix geometries under consideration, although these axioms are not necessarily respected.

\subsection{Lessons and cautions}
\label{sec:lessons}

We draw the following general lessons from the above examples:
\begin{itemize}

\item
The algebra $\cA = \cL(\cH)$ of linear operators on $\cH$
should be viewed as quantization of the algebra of functions 
on a symplectic space $(\cM,\omega)$. 
However as abstract algebra, $\cA$ carries no geometrical information, not even the dimension or the topology of 
the underlying space.  
\item
The geometrical information is encoded in the {\em specific matrices} $X^a$, which 
should be interpreted as quantized embedding functions 
\be
X^a \, \sim \, x^a: \quad \cM \hookrightarrow \R^D .
\ee
They encode the embedding geometry, which is contained e.g. in the 
matrix Laplacian. We will learn below how to extract this more 
directly.
The Poisson or symplectic structure is encoded in their commutation relations.
In this way,
even finite-dimensional matrices can describe various geometries to a high precision.

\item
In some sense, every non-degenerate and ``regular'' fuzzy space 
given by the quantization of a symplectic manifold
locally looks 
like some $\R^{2n}_\theta$. The algebra of functions on
$\R^{2n}_\theta$ is infinite-dimensional only because its volume is infinite:
$\dim(\cH)$  counts the number of quantum cells
via the Bohr-Sommerfeld quantization rule, which is nothing but 
the semi-classical relation \eq{int-trace}.
For example, $\C P^n_N$ can be viewed as a compactification of $\R^{2n}_\theta$.

\end{itemize}

This leads to the idea that generic geometries may be described similarly as 
{\em embedded noncommutative spaces} in matrix models, interpreting the matrices $X^a$ as 
 quantized embedding maps $X^a \sim x^a:\,\,\cM \hookrightarrow \R^D$.
However, we caution that general matrices do not necessarily admit a geometrical interpretation.
There is not even a notion of dimension in general.
In fact we will see that matrix models can describe much more general situations,  
such as multiple submanifolds (''branes``),
intersecting branes, manifolds suspended between
branes, etc., essentially the whole zoo of string theory.
Therefore we have to make some simplifying assumptions, 
and  focus on the simplest case of NC branes 
corresponding to smooth submanifolds.
This will be justified in section \ref{sec:deformations} by showing that such configurations are stable under small deformations.
Moreover, we will explain in section \ref{sec:generic-4D} how to realize 
a large class of generic 4d geometries through such matrix geometries. 

A sharp separation between admissible and non-admissible matrix geometries would in fact be 
inappropriate in the context of matrix model, whose quantization is defined 
in terms of an integral over the space of {\em all} matrices as discussed below.
The ultimate aim is to show that the dominant
contributions to this integral correspond to matrix configurations which have a 
geometrical meaning relevant to physics. However, the integral is over all possible matrices,
including geometries with different dimensions and topologies.
It is therefore clear that such a geometric notion can only be approximate or emergent.

Finally, we wish to address the issue of finite-dimensional versus infinite-dimensional 
matrix algebras.
Imagine that our space-time was fuzzy with a UV scale $\LNC \approx \L_{\rm Planck}$, 
and compact of size $R$. 
Then there would be only finitely many ``quantum cells'', and the geometry should 
be modeled by some finite $N$--dimensional (matrix) algebra. Since no experiment on earth
can directly access the Planck scale, such a scenario can hardly be ruled out, and
a model based on a finite matrix geometry might be perfectly adequate. Therefore the limit
$N \to \infty$ or $\LNC \to \infty$ may not be realized in physics.
However there must be a large 
``separation of scales``, and this limits should be well-behaved
in order to have any predictability; whether or not the limit is realized 
in nature is then irrelevant.

\section{Spectral matrix geometry}

We want to understand more generally such matrix geometries, described by a number 
of hermitian matrices $X^a \in \cA= \cL(\cH)$. Here $\cH$ is a finite-dimensional
or infinite-dimensional (separable) Hilbert space.

One way to extract geometrical information from a space $\cM$ which naturally generalizes to
the noncommutative setting is via spectral geometry. 
In the classical case, one can
consider the heat kernel expansion of the Laplacian $\Delta_g$  of a compact Riemannian manifold $(\cM,g)$
\cite{Gilkey:1995mj},
\be
{\rm Tr} e^{-\a \Delta_g} = \sum_{n\geq0} \a^{(n-d)/2} \int_\cM d^d x \sqrt{|g|}\, a_n(x) .
\label{heatkernel-expand}
\ee
The Seeley-de Witt coefficients $a_n(x)$ of this asymptotic expansion are determined by the intrinsic geometry of $\cM$, 
e.g. $a_2 \sim -\frac{R[g]}6$ where $R[g]$ is the curvature scalar. This provides physically valuable 
information on $\cM$, and describes the one-loop effective action.
In particular, the leading term allows to compute the number of eigenvalues below some cutoff,
\be
\cN_\Delta(\L) := \#\{\mu^2\in {\rm spec} \Delta;\, \mu^2\leq \L^2 \} ,
\ee
dropping the subscript $g$ of the Laplacian.
One obtains Weyls famous asymptotic formula
\be
\cN_\Delta(\L) \sim c_d{\rm vol}\cM \,\L^{d} , \qquad c_d = \frac{{\rm vol} S^{d-1}}{d(2\pi)^d} .
\ee
In particular, the (spectral) dimension $d$ of $\cM$ can be 
extracted the from the asymptotic density of the eigenvalues of $\Delta_g$.
However, although the spectrum of $\Delta_g$ contains a lot of information on the geometry, it 
does not quite determine it uniquely, 
and there are inequivalent but isospectral manifolds\footnote{One way to close this gap is to consider
spectral triples associated to Dirac operators \cite{Connes:1996gi}. 
In the matrix model, the geometrical information will be extracted more 
directly using the symplectic structure and the embedding
defined by the matrices $X^a$.}.

Now consider the spectral geometry of fuzzy spaces in more detail. 
Since the asymptotic density of eigenvalues vanishes in the compact case, 
the proper definition of a spectral dimension in the fuzzy case 
with Laplacian $\Box$ must take into account the cutoff, e.g. as follows 
\be
\cN_{\bDelta}(\L) \sim c_d{\rm vol}\cM \,\L^{d} \qquad\mbox{for}\,\, \L \leq \L_{\rm max}
\ee
where $\L_{\rm max}$ is the (sharp or approximate) cutoff of the spectrum. 
Similarly, the information about the geometry of $\cM$ is encoded in the spectrum of its 
Laplacian or Dirac operator {\em below its cutoff}. 
It turns out that such a cutoff is in fact  
essential to obtain meaningful Seeley-de Witt coefficients in the noncommutative case \cite{Blaschke:2010rr}.
We can thus formulate a specific way to associate an effective geometry to a noncommutative space
with Laplacian $\Box$:
if ${\rm spec}\, \Box$ has a clear enough asymptotics for $\L \leq \L_{\rm max}$
and approximately coincides with ${\rm spec} \Delta_g$ for some classical manifold $(\cM,g)$ for
$\L \leq \L_{\rm max}$, then its spectral geometry is that of $(\cM,g)$.

To proceed, we need to specify a Laplacian for matrix geometries. 
Here the (Yang-Mills) matrix model \eq{MM-action} provides 
a natural choice: For any given background configuration in the matrix model defined by
$D$ hermitian matrices $X^a$, there is a natural
matrix Laplace operator\footnote{This 
operator arises e.g. as equation of motion for the Yang-Mills matrix model.
There is also a natural matrix 
Dirac operator $\slashed{D}\Psi = \Gamma_a \left[X^a, \Psi\right]$ where $\Gamma_a$ generates the Clifford 
algebra of $SO(D)$. However we will not discuss it here.}
\be
{\bDelta} = [X^a,[X^b,.]]\d_{ab}
\label{matrix-laplacian}
\ee
which is a (formally) hermitian operator on $\cA$. We can study its spectrum and 
the distribution of eigenvalues. This Laplacian governs the fluctuations 
in the matrix model, and therefore encodes its effective geometry.
Hence if there is a classical geometry which effectively describes the matrix background $X^a$ up to some scale $\LNC$, 
the spectrum of its canonical (Levi-Civita) Laplacian $\Delta_g$ must approximately coincide 
with the spectrum of $\bDelta$, up to 
some possible cutoff $\L$. In particular,  there should be 
a refined version of the quantization map \eq{quant-map} 
\be
\begin{array}{rcl}
\cI: \quad \cC_{\L}(\cM) &\to& \cA\,\,\subset \,\, \Mat(\infty,\C)\, \\
 f(x) &\mapsto& F
\end{array}
\label{quant-map-2} 
\ee
which approximately intertwines the Laplacians $\cI(\Delta_g f) \approx \bDelta(\cI(f))$.
Here $\cC_\L(\cM)$ denotes the space of functions on $\cM$ whose 
eigenvalues are bounded by $\L$, and $\cI$ should be injective.
The fuzzy sphere is an example where the matrix Laplacian precisely matches the 
classical Laplacian up to the cutoff. Its special symmetry is no longer essential.

\section{Embedded noncommutative spaces and their geometry.}
\label{sec:embedded-branes}

In practice, it is  hard to 
extract information on the metric from the spectrum. A more direct handle on the geometry 
can be obtained for embedded noncommutative spaces, which 
can be understood as quantization of an approximate 
classical symplectic manifold $(\cM,\theta^{\mu\nu})$ embedded in $\R^D$. 
This makes matrix models much more accessible than abstract NC geometry.

Thus consider again matrices $X^a$ which can be viewed as quantized Cartesian embedding functions of $\cM$ in $\R^D$,
\be
X^a \sim x^a: \cM \hookrightarrow \R^D .
\ee
One can then  interpret commutators as quantization of
the Poisson structure on $\cM$. 
In particular,
\be
i\Theta^{ab} \ = \ \, [X^a,X^b] \sim i \{x^a,x^b\} = i \theta^{\mu\nu}\partial_\mu x^a \partial_\nu x^b
\ee
in the semi-classical limit, where $\theta^{\mu\nu}$ is the Poisson tensor in some 
local coordinates on $\cM$.
It is then not hard to see  \cite{Steinacker:2010rh} that
\be
\bDelta \phi \equiv  [X^a,[X^b,\phi]]\d_{ab} \sim  -\{X^a,\{X^b,\phi\}\}\d_{ab} = - e^\sigma \Delta_{G} \phi(x) 
\label{laplace-semiclass}
\ee
for any matrix resp. function $\cA \ni \phi \sim \phi(x)$. 
Here $\Delta_{G}$ is the standard Laplace operator associated to the effective metric $G_{\mu\nu}$
defined as follows
\bea  
G^{\mu\nu}(x) &:=& e^{-\sigma}\,\theta^{\mu\mu'}(x) \theta^{\nu\nu'}(x) 
 g_{\mu'\nu'}(x)  
\label{G-def-general}  \\
g_{\mu\nu}(x) &:=& \partial_\mu x^a \partial_\nu x^b \d_{ab} \,\, ,
\label{g-explicit}\\
e^{-(n-1)\sigma} &:=& \frac 1{\theta^{n}}\, |g_{\mu\nu}(x)|^{-\frac 12},
\qquad \theta^n = |\theta^{\mu\nu}|^{1/2} .
\label{sigma-rho-relation} 
\eea
All of these are tensorial objects on $\cM$, e.g.
 $g_{\mu\nu}(x)$ is the metric induced on $\cM\subset \R^{D}$ via 
pull-back of $\d_{ab}$. The normalization factor 
$e^{-\sigma}$ is determined uniquely (except for $n=1$) such that
\be
\frac 1{\theta^{n}} = \sqrt{|G_{\mu\nu}|}\, e^{-\sigma} .
\label{rho-sigma-det}
\ee 
This provides the desired explicit 
description of the matrix geometry at the semi-classical level. 
It is easy to check that $G_{\mu\nu} = g_{\mu\nu}$ for the examples in section \ref{sec:examples},
which will be understood on more general grounds below.

The easiest way to see \eq{laplace-semiclass}
is by considering the action for a scalar field coupled to the matrix model background
\bea
S[\varphi] &\equiv& - {\rm Tr} [X^a,\phi][X^b,\phi] \d_{ab} 
\sim \frac{1}{(2\pi)^n}\, \int d^{2n} x\; 
\sqrt{|G_{\mu\nu}|}\,G^{\mu\nu}(x)
 \partial_{\mu} \phi \partial_{\nu} \phi \,.
\label{covariant-action-scalar}
\eea
Writing the lhs as ${\rm Tr}\phi \Box\phi$ and taking the semi-classical limit
leads to \eq{laplace-semiclass}. 
Note that this is the action for additional matrix components
$\phi \equiv X^{D+1}$ in the matrix model \eq{MM-action}. Therefore 
$G_{\mu\nu}$ is the metric which governs scalar fields
in the matrix model, more precisely nonabelian scalar fields which arise as transversal  fluctuations  
on backgrounds $X^a\otimes \one_n$, cf. \eq{coinciding-branes}.
More generally, one can show that all fields which arise in the matrix model as fluctuations 
of the matrices around such a background (i.e. scalar fields, gauge fields and fermions) are governed by $G_{\mu\nu}$,
possibly up to a conformal factor $\sim e^\sigma$.
This means  that $G_{\mu\nu}$ is the effective gravitational metric.

Now consider the equations of motion of the matrix model \eq{MM-action}, which are given by 
\be
0 = [X^b,[X^{b'},X^a]]g_{bb'} \ =\ \Box X^a \ \sim \  -e^\sigma\, \Delta_G x^a .
\ee
This means that the embedding functions are harmonic functions w.r.t. $G_{\mu\nu}$,
which is  satisfied e.g. for $\R^{2n}_\theta$. With a little more
effort, one can also derive the following equations for the 
Poisson structure \cite{Steinacker:2008ya,Steinacker:2010rh}
\be
\nabla^\mu_G (e^{\sigma}\theta^{-1}_{\mu \nu})
=  e^{-\sigma} \,G_{\nu\rho}\,\theta^{\rho \mu} \partial_\mu \eta, \qquad \eta = \frac 14 e^\sigma G^{\mu\nu} g_{\mu\nu} .
\label{theta-covar-id-text}
\ee
However, quantum corrections are expected to be essential for the gravity sector,
and one should be careful avoiding preliminary conclusions. 
In any case, we note the following observations:

\begin{itemize}
 \item 
Assume that $\dim \cM = 4$. Then $G_{\mu\nu} = g_{\mu\nu}$ if and only if the symplectic form
$\omega = \frac 12 \theta^{-1}_{\mu\nu} dx^\mu dx^\nu$ 
is self-dual or anti-selfdual \cite{Steinacker:2010rh}. 

\item
There is a natural tensor
\be
\cJ^{\eta}_\gamma = e^{-\sigma/2}\, \theta^{\eta\gamma'} g_{\gamma' \gamma}
 = - e^{\sigma/2}\,  G^{\eta \gamma'} \theta^{-1}_{\gamma' \gamma} .
\label{J-tensor}
\ee
Then the effective metric can be written as
\be
G^{\mu\nu} 
= \cJ^{\mu}_\rho\, \cJ^{\nu}_{\rho'}\, g^{\rho\rho'}
= - (\cJ^2)^{\mu}_\rho\, g^{\rho\nu}.
\ee
In particular, $\cJ$ defines an almost-complex structure if and only if $G_{\mu\nu} = g_{\mu\nu}$,
hence for  (anti-)selfdual $\omega$. In that case, 
$(\cM,\tilde g,\omega)$ defines an almost-K\"ahler structure on $\cM$ where
\be
\tilde g_{\mu\nu} := e^{-\sigma/2}\,  g_{\mu\nu} .
\label{almost-K}
\ee 
\item
The matrix model is invariant under 
gauge transformations $X^a \to {X^a}' = U^{-1} X^a U$, which semi-classically 
correspond to symplectomorphisms $\Psi_U$ on $(\cM,\omega)$.
This can be viewed in terms of modified embeddings ${x^a}' = x^a \circ \Psi_U: \,\, \cM \to \R^D$
with equivalent geometry.
\item
Matrix expressions such as $[X^a,X^b] \sim i \theta^{\mu\nu}\del_\mu x^a \del_\nu x^b$ 
should be viewed as (quantizations of) tensorial objects on $\cM\subset \R^D$,
written in terms of Cartesian coordinates $a,b$ of the ambient space $\R^D$.
They are always tangential because $\del_\nu x^b \in T_p\cM$.
Using appropriate projectors on the tangential and normal bundles of $\cM$, this can be 
used to derive matrix expressions which encode e.g. the intrinsic curvature of $\cM$,
cf. \cite{Blaschke:2010rg,Arnlind:2010kw}. This is  important for gravity.

\end{itemize}

\section{Realization of  generic 4D geometries in matrix models}
\label{sec:generic-4D}

We now show how a large class of generic (Euclidean, for now) 
4-dimensional geometries can be realized as NC
branes in matrix models with $D=10$. This should eliminate any lingering doubts whether the geometries 
in the matrix model are sufficiently general for  gravity. This construction is illustrated
in \cite{Blaschke:2010ye} for the example of the Schwarzschild geometry.

\begin{enumerate}
 \item 
Consider some ''reasonable'' generic geometry $(\cM^4,g_{\mu\nu})$ with nice properties, as explained below.
\item
Choose an embedding $\cM \hookrightarrow \R^D$. This is in general not unique, and 
requires that $D$ is sufficiently large. Using classical embedding theorems \cite{clarke}, 
$D=10$ is just enough to embed generic  4-dimensional geometries
(at least locally).

\item
Equip $\cM$ with an (anti-)selfdual closed 2-form $\omega$. This means that
$d\omega = d\star_g \omega = 0$, hence $\omega$ is a special solution of the free Maxwell equations on $\cM$.
Such a solution generically exists for mild assumptions on $\cM$, for example by solving the 
corresponding boundary value problem with 
$\omega$ being (anti-)selfdual on the boundary or asymptotically\footnote{It may happen that 
$\omega$ vanishes at certain locations, cf. \cite{Blaschke:2010ye}. 
This might be cured through compact extra dimensions.}. This defines the requirements in step 1).
For asymptotically flat spaces, $\omega$ should be asymptotically constant in order to ensure
that  $e^{-\sigma}$ is asymptotically constant. This requirement can be met more easily in the 
presence of compact extra dimensions $\cM^4 \times \cK$ \cite{kuntner}.

As explained above, it follows that $(\tilde g,\omega)$ \eq{almost-K} is almost-K\"ahler. 
Under mild assumptions, one can then show 
\cite{uribe} that there exists a quantization \eq{quant-map} of the symplectic space $(\cM,\omega)$ in terms of 
operators on a Hilbert space\footnote{The use of the almost-K\"ahler structure 
may only be technical and should actually not be necessary.}.
In particular, we can then define $X^a := \cI(x^a)\in \cA$ to be the matrix obtained as quantization of $x^a$,
so that
\be
X^a \sim x^a: \cM \to \R^D .
\ee
The effective metric on $\cM$ is therefore $G_{\mu\nu}$ as explained above.
\item
Since $\omega$ is (anti-)selfdual it follows that $G = g$, and we have  indeed obtained a quantization of $(\cM,g)$
in terms of a matrix geometry. In particular, the matrix Laplacian $\bDelta$ will approximate $\Delta_g$ for low enough
eigenvalues, and fluctuations of the matrix model around this background
describe fields propagating on this effective geometry. 

\end{enumerate}

\section{Deformations of embedded NC spaces}
\label{sec:deformations}

Assume that $X^a \sim x^a:\,\cM \hookrightarrow \R^D$ describes some quantized embedded space as before.
The important point which justifies the significance of this class of configurations
is that {\em it is preserved by small deformations}. 
Indeed, consider a small deformation $\tilde X^a =  X^a +A^a$ by generic matrices $A^a \in \cA$.
By assumption, there is a local neighborhood 
for any point $p\in \cM$ where we can separate the matrices $X^a$ into 
independent coordinates and embedding functions,
\be
X^a = (X^\mu,\phi^i(X^\mu)) 
\label{X-splitting}
\ee
such that the $X^\mu$ generate the full\footnote{In topologically non-trivial situations 
they will individually generate only ``almost`` the full $\cA$, and $\cA$
is recovered by combining various such local descriptions. This will become more clear in the example of $S^2_N$.} 
matrix algebra $\cA$.
Therefore we can write  $A^a = A^a(X^\mu)$, and assume that it is smooth
(otherwise the deformation will be suppressed by the action). 
We can now consider $\tilde X^\mu = X^\mu+A^\mu \sim \tilde x^\mu(x^\nu)$ as new coordinates with modified 
Poisson structure $[\tilde X^\mu,\tilde X^\nu] \sim i \{\tilde x^\mu,\tilde x^\nu\}$, and 
 $\tilde \phi^i = \phi^i + A^i \sim \tilde \phi^i(\tilde x^\mu)$ as modified embedding 
of $\tilde \cM \hookrightarrow \R^D$. Therefore $\tilde X^a$ describes again a quantized embedded space.
Due to this stability property, it is plausible that the class of embedded NC spaces plays a dominant role in the 
path integral \eq{Z-def}.

To obtain an intuition and to understand the local description, 
consider the example of the fuzzy sphere. 
We can solve for $X^3 = \pm \sqrt{1 - (X^1)^2 - (X^2)^2}$, and use 
$X^1, X^2$ as local coordinate ``near the north pole'' $X^3 = +1$ or the south pole $X^3 = -1$.
Each branch of the solution makes sense provided some restriction on the spectrum of 
$X^3$ is imposed, and in general ``locality`` might be phrased as a condition on the spectrum
of some coordinate(s).
Then the $X^1,\, X^2$ ''locally'' generate the full matrix algebra $\cA$.

The splitting \eq{X-splitting} can be refined using 
the  $ISO(D)$ symmetry of Yang-Mills matrix models. In the semi-classical limit,
one can assume (after a suitable rotation) that $\partial_\mu \phi^i = 0$ 
for any given point $p \in \cM$, so that the tangent space
is spanned by the first $d$ coordinates in $\R^D$. Moreover, $p$ can be moved to the origin using the 
$D$-dimensional translations.
Then the matrix geometry looks locally  
like $\R^d_\theta$, which is deformed geometrically by non-trivial $\phi^i(X^\mu)$ and a non-trivial
commutator $[X^\mu,X^\nu] = i (\bar \theta^{\mu\nu} + \d \theta^{\mu\nu}(X^\a))$. 
These $X^\mu\sim x^\mu$ define ''local embedding coordinates``,
which are analogous to Riemannian normal coordinates.
Hence any deformation of 
$\R^d_\theta$ gives a matrix geometry as considered here, and vice versa any 
"locally smooth'' matrix geometry  should have such a local description. 
This justifies our treatment of matrix geometry.


\section{Generalized backgrounds}

Although we focused so far on matrix geometries which are quantizations of classical 
symplectic manifolds, 
matrix models are much richer and accommodate 
structures such as multiple branes,
intersecting branes, manifolds suspended between branes, etc. 
For example, consider the following block-matrix configurations\footnote{
Recall that by the Wedderburn theorem, the algebra generated by (finite-dimensional) 
hermitian matrices is always a product of simple matrix algebras,
i.e. it decomposes into diagonal blocks as in \eq{block-branes}.}
\be
Y^a = \left(\begin{array}{cccc}
                           X^a_{(1)} & 0 & 0 & 0 \\
                           0 & X^a_{(2)} & 0 & 0 \\
                           0 & 0 & \ddots & 0 \\
                           0 & 0 & 0 &  X^a_{(n)}
                           \end{array}\right) .
\label{block-branes}
\ee
Assuming that each block $X^a_{(i)}$ generates the (matrix) algebra $\cA_{(i)}$ of functions 
on some quantized symplectic space $\cM_{(i)}\subset \R^D$, it is clear that the matrices $Y^a$ should be interpreted as 
$n$ different NC branes embedded in $\R^D$. One way to see this is by considering optimally localized 
(''coherent``, cf. section \ref{sec:coherent-states}) states $|x\rangle_{(i)}$ corresponding to each block, 
such that the VEVs
${}_{(i)}\langle x | X^a | x \rangle_{(i)} \approx x^a$ approximately sweep the location of $\cM_{(i)}\subset \R^D$.
Then clearly these state are also optimally localized for $Y^a$, and together 
sweep out the multiple brane configuration\footnote{Elements of the off-diagonal blocks are
naturally identified as bi-modules or ''strings`` connecting the branes. 
However we will not consider this here.}. 
One particularly important case is given by  a 
stack of $n$  coinciding branes:
\be
 Y^a = X^a \otimes \one_n \, .
\label{coinciding-branes}
\ee
Fluctuations around such a background lead to $SU(n)$ Yang-Mills gauge theory on $\cM$,
as explained below. In fact, 
the underlying algebra $\cA \otimes \Mat(n,\C)$ can be interpreted in
two apparently different but nonetheless equivalent ways: 1) as $su(n)$ valued functions on $\cM$
or 2) describing a higher-dimensional space $\cM \times \cK$, where $\Mat(n,\C)$ is 
interpreted as quantization of some compact symplectic space $\cK$.
Which of these two interpretations is physically correct depends on the actual 
matrix configuration, generalizing \eq{coinciding-branes}. Such extra dimensions allow to add more 
structure such as physically relevant gauge groups etc., 
cf.  \cite{Aschieri:2006uw}. 

Another interesting case if that of intersecting branes, such as two Moyal-Weyl quantum planes embedded along
different directions. This is very useful to make contact 
with particle physics \cite{Chatzistavrakidis:2011gs}, but
we only mention it here to illustrate the rich zoo of backgrounds of the matrix model, which is not
restricted to smooth geometries.

\section{Effective gauge theory}

Consider a small deformation of the Moyal-Weyl quantum plane embedded in $\R^{D}$, 
\be
X^a = \begin{pmatrix}
 \bar X^\mu \\ 0
\end{pmatrix} \, + \, 
\begin{pmatrix}
 -\bar\theta^{\mu\nu} A_\nu  \\   \phi^i
\end{pmatrix} .
\label{eq:general-X}
\ee
This can be interpreted either in terms of  deformed geometry explained above, or
in terms of NC gauge theory  by
considering $A_\mu = A_\mu(\bar X)$ and $\phi^i = \phi^i(\bar X)$ as gauge fields and scalar
fields on $\R^4_\theta$. 
More precisely, fluctuations around a stack of branes \eq{coinciding-branes}
turn out to describe $\msu(n)$-valued gauge fields
coupled to the effective metric $G_{\mu\nu}$, and the matrix model \eq{MM-action}
yields the effective action \cite{Steinacker:2010rh}
\bea
S_{YM}[A] &\sim&
\frac{\Lambda_0^4}{4}\int d^{4} x\,  e^{\sigma} \, \sqrt{|G_{\mu\nu}|}
G^{\mu\mu'} G^{\nu\nu'} 
{\rm tr}(F_{\mu\nu}\,F_{\mu'\nu'})\,\, 
+\,\, \frac 12 \int \eta F\wedge F 
\label{action-expanded-2}
\eea
(dropping $\phi^i$).
On $\R^4_\theta$, this is easy to understand:
the gauge transformations $X^a \to U X^a U^{-1}$ give rise to
$$
A_\mu \to U  A_\mu U^{-1}+i\ U \partial_\mu U^{-1}
$$
using \eq{eq:general-X}, and
the field strength $F_{\mu\nu}= \partial_\mu A_\nu -  \partial_\nu A_\mu + i[A_\mu,A_\nu]$ 
is encoded in the  commutator,
$$
\,[X^\mu,X^\nu] = -i \bar\theta^{\mu\mu'} \bar\theta^{\nu\nu'}
(\bar\theta^{-1}_{\mu'\nu'} + F_{\mu'\nu'}).
$$ 
Then \eq{action-expanded-2} follows up to surface terms.
However, the trace-$U(1)$ components of $A_\mu$ and $\phi^i$ should really be interpreted
as embedding fluctuations of the brane, defining the effective metric
$G_{\mu\nu}$ on a general $\cM \subset \R^D$. Then the derivation of \eq{action-expanded-2}
becomes somewhat more technical, see \cite{Steinacker:2007dq,Steinacker:2008ya}. 
Nevertheless the gauge theory point of view is useful to carry out the quantization,
as discussed below. This is a key feature of 
NC emergent gravity which greatly simplifies its quantization.

\section{Quantization and effective action}
\label{sec:quantization}

Up to now, we discussed the geometry of some given NC space or brane in matrix models. 
It should be clear that such a quantum geometry does not amount to 
the quantization of a physical theory; it is a deformed classical geometry. 
To talk about quantum field theory or quantum gravity, we need to quantize the degrees of freedom
in the  model. 

There are various ways of quantizing a classical theory: one can 
follow canonical quantization via a phase space formulation of the classical model, or
attempt some sort of path-integral  quantization. 
For matrix models, there is a very natural approach which is the analog of configuration-space
path integral: Quantization is defined as an integration over the space of matrices. 
More explicitly, the partition function is defined as
\be
Z[J] = \int dX^a\, e^{-S[X] + X^a J_a} ,
\label{Z-def}
\ee
where we introduced external matrices $J^a$ in order to compute correlation functions 
and the effective action. The extension to fermions is straightforward. 
This definition respects the fundamental gauge symmetry $X^a \to U X^a U^{-1}$, as well
as all global symmetries (translations, rotations, and possibly SUSY). 
These matrix integrals are known to 
exist\footnote{more precisely, this has been established for the partition function 
and certain correlation functions \cite{Austing:2001pk}.} 
 for finite $N$.

Since the matrices $X^a$ encode both the geometry of the branes and also the propagating fields such as 
gauge fields, this matrix integral comprises quantum field theory as well as a quantum theory of 
geometry, hence of gravity. Moreover, there is no way
to separate the field theoretical from the geometrical degrees of freedom. Hence we are 
facing a unified quantum theory of fundamental interactions including gravity, or some toy model thereof.

To obtain a potentially realistic model which describes near-classical geometries, 
the limit $N\to\infty$ of the quantization \eq{Z-def} must exist in some sense. 
This is of course  highly non-trivial,
and issues such  UV-divergences, renormalization etc. arise in the $N\to\infty$ limit. 
In fact for most matrix models this limit probably does not make sense, 
or is very different from the semi-classical picture.
However, there is essentially a {\em unique} model within this class of models (with $D\geq 4$)
where this limit can be expected to exist, due to its maximal supersymmetry:
the IKKT model \cite{Ishibashi:1996xs}
\be
S_{\rm IKKT} =-(2\pi)^2{\rm Tr}\Big(\,[X^a,X^b][X_a,X_b]\,\, + \, 2\overline\Psi \gamma_a[X^a,\Psi] \Big) 
\,,
\label{IKKT-MM}
\ee
where $D=10$ and $\Psi$ are Majorana-Weyl spinors of $SO(9,1)$. On 4-dimensional NC brane backgrounds 
$\R^4_\theta$, this model can be viewed  as $\cN=4$ NC super-Yang-Mills gauge theory on $\R^4_\theta$,
which is expected to be finite (at least perturbatively, but arguably also beyond)
just like its commutative version. We will discuss some pertinent points 
below, and establish in particular one-loop finiteness. 

A remark on the signature is in order. 
Majorana-Weyl spinors in $D= 10$ exist only for Minkowski signature, transforming under the 10-dimensional
Lorentz group $SO(9,1)$. This is of course the physically relevant case, 
and accordingly there should be an $i$ in the exponent in \eq{Z-def}.
However the Euclidean model does make sense e.g. after integrating out the fermions, and is mathematically
easier to handle. We will therefore  continue our discussion in the Euclidean case.


\subsection{UV/IR mixing in noncommutative gauge theory}

Using the gauge theory point of view, the quantization of matrix fluctuations  in \eq{eq:general-X}
can be carried out using standard field theory techniques adapted to the NC case.
We need to quantize the gauge fields $A_\mu = A_\mu(\bar X)$ and scalar fields 
$\phi^i = \phi^i(\bar X)$ on $\R^4_\theta$. 
A general function on $\R^4_\theta$ can be expanded in
a basis of plane waves, e.g.
\be
\phi(X) = \int \frac{d^4 k}{(2\pi)^4}\, \phi_k \, e^{i k_\mu \bar X^\mu}
\quad \in\,\, \Mat(\infty,\C) \cong \R^4_\theta
\ee
where $\phi_k$ is an {\em ordinary} function of $k \in \R^4$. 
Inserting this into the action \eq{MM-action},
the free (quadratic) part is then independent of
$\theta^{\mu\nu}$, but the interaction vertices 
acquire a nontrivial phase factor 
$e^{\frac i2 \sum_{i<j} k^i_{\mu}k^j_{\nu}\theta^{\mu\nu}}$
where $k^i_{\mu}$ denotes the incoming momenta.
The matrix integral  becomes an ordinary integral 
$\int d X^a = \int \Pi d \phi_k$,
which can be evaluated perturbatively in terms of 
Gaussian integrals. Thus Wicks theorem follows, however planar and 
non-planar contractions are distinct due to these phase factors, leading to  Feynman-Filk rules.
It turns out that planar diagrams coincide with their 
undeformed counterparts, while the non-planar diagrams 
involve oscillatory factors. 

One can now compute correlation functions and loop contributions. In particular, 
the planar loop integrals have the same divergences as in the commutative case, 
in spite of the existence of a fundamental scale $\L_{NC}$.
This should not be too surprising, because noncommutativity leads to a quantization of area but not of length. 
However, a (virtual) UV momentum in some direction $k \gg \L_{NC}$ necessarily implies a non-classical 
IR effect in another direction. 
This leads to the infamous UV/IR mixing \cite{Minwalla:1999px}, which technically originates 
from oscillatory integral due to the phase factors in non-planar diagrams. 
This phenomenon is ubiquitous in NC field theory, and  leads to pathological 
IR divergences in any UV-divergent model, which cannot be cured by standard 
renormalization\footnote{Renormalizable models do exist \cite{Grosse:2004yu}, at the 
expense of modifying the infrared behavior of the model e.g. through a confining potential.
We refer to the contribution by H. Grosse and M. Buric in this volume.}.

As an illustration, we display the ''strange`` contribution of a scalar fields coupled to some external $U(1)$ gauge field
in the 1-loop effective action:
\bea
\Gamma_{\Phi} 
&=& - \frac{g^2}2\frac{1}{16 \pi^2}\,\int \frac{d^4 p}{(2\pi)^4}\, \, 
 \Big(-\frac 1{6} F_{\mu\nu}(p) F_{\mu'\nu'}(-p) 
 g^{\mu\mu'} g^{\nu\nu'}
 \,\log(\frac{\L^2}{\L_{\rm eff}^2}) \nn\\
&& + \frac 14 (\theta F(p)) (\theta F(-p))
 \Big(\L_{\rm eff}^4 - \frac 16 p\cdot p\, \L_{\rm eff}^2 
 +\frac{(p\cdot p)^2}{1800}\, 
(47-30\log({\textstyle\frac{p\cdot p}{\L_{\rm eff}^2}})) 
 \Big)\Big)  
\label{induced-terms-UVIR}
\eea
where
\be
\frac 1{\L_{\rm eff}^2(p)} = 
 \frac 1{\Lambda^2} + \frac 14 \frac{p^2}{\L_{NC}^4} .
\nn
\ee
is finite for $p \neq 0$ but diverges for $p \to 0$ and $\L \to \infty$. 
These IR divergences become worse in higher loops,
and the models are probably pathological as they stand.

For NC gauge theories as defined by matrix models, the geometrical insights explained
above allows to understand this phenomenon 
in physical terms: Since fluctuations in the matrix model are understood as fields coupled to 
a non-trivial background metric, it follows that their quantization necessarily leads to induced gravity 
action, which diverge as $\L \to \infty$. This is the standard mechanism of induced gravity due to Sakharov.
This explanation of UV/IR mixing holds in the semi-classical regime i.e. for low enough cutoff, and has been 
verified in detail \cite{Blaschke:2010rr,Grosse:2008xr}
that these induced gravity terms give e.g. \eq{induced-terms-UVIR} in the appropriate limit.

We can now turn this problem into a virtue, noting that there is essentially one unique 
matrix model which does not have this problem (in 4 dimensions), given by the $\cN=4$ SYM theory on $\R^4_\theta$,
or equivalently the IKKT model \eq{IKKT-MM}.
This is (almost) the unique model which is arguably well-defined
and {\em UV finite} to any order in perturbation theory, 
hence no such IR divergences arise\footnote{These are not rigorous results at present 
but well justified by the relation with the commutative model \cite{Jack:2001cr}, and partially verified by some loop 
computations in the NC case.}.


Accepting finiteness in the gauge theory point of view, it follows immediately from our
geometrical discussions that the model provides a well-defined quantum theory of dynamical geometry in 4 dimensions,
hence of (some type of) gravity. Moreover, there are clearly relations with general relativity:
There are induced Einstein-Hilbert terms in the quantum effective action 
due to \eq{heatkernel-expand}, and moreover
on-shell fluctuations of the would-be $U(1)$ gauge fields have been shown to be Ricci-flat 
metric fluctuations \cite{Rivelles:2002ez}. Nevertheless, at present there is no satisfactory understanding of the 
dynamics of this emergent gravity, due to the complexity of the system involving 
the Poisson structure. The presence of compactified extra dimensions can also be expected to play an 
important role here, and more work is needed to understand the effective gravity in this model.

\subsection{1-loop quantization of IKKT model}

In this last section, we illustrate the power and simplicity of the model by computing the full 1-loop effective action, 
and establish that it is UV finite on $\R^4_\theta$.
The action induced by integrating out the 
fermions\footnote{there is a subtlety -- a Wess-Zumino contribution is missing here.} is 
given as usual by 
$\Gamma_{{\rm 1-loop}}^{(\Psi)} = -\frac 14 \Tr \log \slashed{D}^2 
= -\frac 14 \Tr\log(\Box  + \Sigma^{(\psi)}_{ab}[\Theta^{ab},.])$, 
where
\begin{align}
(\Sigma_{ab}^{(\psi)})^\a_\b &\ =\  \frac i4 \,[\gamma_a,\gamma_b]^\a_\b \qquad \qquad\ \ \mbox{fermions}  \\
(\Sigma_{ab}^{(Y)})^c_d &\ =\  i(\d^c_a g_{bd} - \d^c_b g_{ad}) \qquad \mbox{bosonic matrices}
\end{align}
denote the generators of $SO(10)$ on the spinor and vector representations.
To quantize the bosonic degrees of freedom is slightly more tricky due to the gauge invariance. 
We can use the background field method, splitting the matrices into background $X^a$ and a fluctuating part $Y^a$, 
\be
X^a\to X^a+Y^a . 
\ee
For a given background $X^a$, the gauge symmetry becomes
 $Y^a \to  Y^a + U [X^a + Y^a, U^{-1}]$,
which we fix using the gauge-fixing function $G[Y] = i[X^a,Y_a]$. This can be done as usual
using the Faddeev-Popov method or alternatively using BRST \cite{Blaschke:2011qu}. 
Then the one-loop effective action induced by the bosonic matrices $Y^a$ is obtained  as 
$\Gamma_{{\rm 1-loop}}^{(Y)} = \frac 12 Tr \log(\Box  +  \Sigma^{(Y)}_{rs}[\Theta^{rs},.]) - Tr \log(\Box)$,
where the last term is due to the FP ghosts. Hence
the full contribution for the IKKT model is given by \cite{Ishibashi:1996xs,Blaschke:2011qu}
\begin{align}
\Gamma_{\!\textrm{1loop}}[X]\! 
&= \frac 12 \Tr \left(\log(\Box  + \Sigma^{(Y)}_{ab}[\Theta^{ab},.])
-\frac 12 \log(\Box  + \Sigma^{(\psi)}_{ab}[\Theta^{ab},.])
- 2 \log \Box\right)   \nn\\
&= \frac 12 \Tr \left(\log(\one  + \Sigma^{(Y)}_{ab}\Box^{-1}[\Theta^{ab},.])
-\frac 12 \Big(\log(\one  + \Sigma^{(\psi)}_{ab}\Box^{-1}[\Theta^{ab},.])\right) \nn\\
 &= \frac 12 \Tr \Bigg( -\frac 14 (\Sigma^{(Y)}_{ab} \Box^{-1}[\Theta^{ab},.] )^4 
 +\frac 18 (\Sigma^{(\psi)}_{ab} \Box^{-1}[\Theta^{ab},.])^4 \,\, +  \cO(\Box^{-1}[\Theta^{ab},.])^5  \Bigg) .
\end{align}
The first 3 terms in this expansion cancel, which reflects the maximal supersymmetry. 
The traces are clearly UV convergent on 4-dimensional backgrounds such as $\R^4_\theta$, so that 
the 1-loop effective action is well-defined. Note that it incorporates both gauge fields and scalars, 
hence all gravitational degrees of freedom from the geometric point of view; 
for a more detailed discussion see \cite{Blaschke:2011qu}. 
Finiteness only holds for the IKKT model, while for all 
other models  of this class with $D\neq 10$ already this 1-loop action is divergent. 
These divergences are in fact much more
problematic than in the commutative case and cannot be handled with  standard renormalization techniques, due  to 
UV/IR mixing. Hence the NC case is much more selective than the commutative case, and the existence of an 
essentially unique well-behaved model is very remarkable.

Another remarkable aspect of this result is that the global $SO(10)$ invariance is manifestly preserved, 
and broken only {\em spontaneously} through the background brane such as $\R^4_\theta$.
Such a statement would be out of reach within conventional quantum field theory, and
 noncommutativity is seen to provide remarkable new tools and insights.

At higher loops, perturbative finiteness is expected, because the UV divergences are essentially the same as in
commutative $\cN=4$ SYM \cite{Jack:2001cr}. Moreover UV/IR mixing results from the divergences at
higher genus, which should also vanish by appealing to the large $N$ expansion of the commutative model.
Alternatively, 1-loop finiteness along with $\cN=1$ supersymmetry and the global  
$SO(10)$ or $SO(9,1)$ symmetry should ensure perturbative finiteness. These arguments remain to be made precise.

\section{Further aspects and perspectives}

Since the cosmological constant problem was raised in the introduction, we should briefly comment on this
issue. Given our very limited understanding at present, no serious claims can be made. However, 
there are  intriguing observations which raise the hope that this problem might be resolved here.
The main point is that the metric is not a fundamental geometrical degrees of freedom, but a 
composite object which combines both the embedding of the brane $\cM^4 \subset \R^{D}$ and its 
Poisson tensor. This means that the equations of motion are fundamentally different from the Einstein equations even
if the effective action has the standard Einstein-Hilbert form, and there will be new types of
solutions which are less sensitive to the vacuum energy \cite{Steinacker:2010rh}.

These different degrees of freedom are also the reason why 
the IKKT matrix model can be perturbatively finite, unlike  general relativity. 
This  model is much more suitable for
quantization than GR. However, it remains to be shown that it also provides a physically viable 
description of gravity. There are several indications which suggest that this should be the case,
including  Ricci-flat deformations on Moyal-Weyl space \cite{Rivelles:2002ez}, 
the relation with IIB supergravity and string theory \cite{Ishibashi:1996xs},
the possibility to obtain Newtonian gravity \cite{Steinacker:2009mp},  
the fact that it gives {\em some} gravity theory with sufficiently 
rich class of geometries, etc.. However, the complicated interplay  of the various degrees of freedom
is not yet well understood, and more work is required before conclusions on the physical viability 
of this approach to quantum gravity can be drawn.

Finally a comparison with string theory is in order, since the IKKT model was proposed originally as a 
non-perturbative definition of IIB string theory. 
The  link with IIB  supergravity and string theory is established only for the interactions between brane
solutions of the type we considered here. 
However, it may not reproduce e.g. all the massive degrees of freedom in string theory.
Hence it seems more appropriate to consider the matrix model as a spin-off from string theory, which 
does provide a good quantum theory of 3+1-dimensional branes, but not necessarily for full string theory. In fact the 
1-loop action is ill-defined e.g. on 8- and 10-dimensional branes, and the degrees of freedom of the metric 
are not fundamental but emergent and composite\footnote{There is no problem with the Weinberg-Witten theorem
which applies only for Lorentz-invariant field theories.}. This in turn allows to avoid many problems of 
string theory, notably the lack of predictivity as illustrated in the landscape issue, 
while preserving many of its attractive features in a simpler framework.

In any case, this and related matrix models
provide exciting new candidates for a quantum theory of gravity coupled to matter, and certainly deserve a 
thorough investigation.

\subsection*{Acknowledgments}

I would like to thank the organizers of the $3^{rd}$ Quantum Gravity and Quantum Geometry School in Zakopane 2011  
for providing a pleasant venue for stimulating and lively discussions with participants from various backgrounds.    
This work was supported by the Austrian Science Fund (FWF), project P21610.

\end{document}